\documentclass[preprint,pre,aps]{revtex4}
\usepackage[dvips]{graphicx}
\bibstyle{prsty}
\begin{document}
\title{Simple theory for oscillatory charge profile in ionic liquids near a charged wall}
\author{ A. Ciach}
\address{Institute of Physical Chemistry,
 Polish Academy of Sciences, 01-224 Warszawa, Poland}
 \date{\today} 
 \begin{abstract}
  The mesoscopic field theory for ionic systems [A. Ciach and G. Stell, J. Mol. Liq. {\bf 87}, 255 (2000)] 
  is extended to the system with charged boundaries. 
  A very simple expression for the excess grand potential functional
  of the charge density is developed.  The size of hard-cores of ions is taken into account in the 
  expression for the internal energy. The functional is suitable
  for a description of a distribution of ions in ionic liquids 
  and ionic liquid mixtures with neutral components near a  weakly  charged wall. The
  Euler-Lagrange equation is obtained, and solved for a 
  flat  confining surface. An exponentially damped  oscillatory 
  charge density profile is obtained. The  electrostatic potential for the restricted primitive model 
  agrees  with the simulation results on a semiquantitative level. 
 
 \end{abstract}

 \maketitle
\section{Introduction}

Ionic systems have been studied for decades. At the beginning, the main interest concerned diluted systems or
electrolytes. The theory of such systems is very well developed, and one of the main contributors was
Lesser Blum. His works on the mean spherical approximation (MSA) and its various extensions applied to
ionic solutions, especially Refs.~\cite{blum:75:0,blum:77:0},
have been very influential. 
Apart from the bulk properties, studies focused on the distribution of ions near a charged wall. 
In the works~\cite{henderson:78:0,blum:81:0} by Douglas Henderson and Lesser Blum, exact relation for the contact value of the wall-fluid 
correlation function,
and exact asymptotic expression for large distances from the wall of this function are given; 
in addition, MSA results for this system are presented.  The liquid-matter theories, however, are rather difficult.

Properties of electrolytes near a charged surface are very well described by the elegant and simple theories of
the mean-field (MF) type, in particular by  the 
Debye-Hueckel (DH) theory~\cite{barrat:03:0,israel:11:0,fedorov:14:0}. Ions in these theories are
treated as point-like charges. In the case of ionic liquids, however, these theories cannot
correctly predict the oscillatory decay of the charge density as the separation from the weakly charged 
surface increases~\cite{lynden-bell:03:0,lynden-bell:05:0,iwahashi:09:0,fedorov:14:0,fedorov:08:0}. It is commonly
assumed that the monotonic decay of the charge density results from the MF character of the theories. 
However, for large densities of the ions the finite size of their cores starts to play an important role.
 The  hard-core packing is carefully taken 
into account in the 
density functional  theory (DFT)~\cite{evans:79:0} that is of the MF type. 
The DFT correctly predicts the oscillatory decay of the density profile
in neutral fluids confined by solid walls~\cite{roth:02:0}. An oscillatory decay of the charge density  near
a confining surface is a natural consequence of the  oscillatory decay of the densities of the 
anions and the cations. A role similar to  steric repulsions is played by the  forbidden multiple occupancy of lattice 
sites in lattice models. As a consequence,
 damped oscillations of the charge ware obtained in a one-dimensional lattice Coulomb gas by exact calculations~\cite{demery:12:0}. 

The decay of the excess density near a wall mimics the decay of the density-density correlation function 
in the bulk, up to the amplitude and phase depending on the properties of the surface.
In the  theory 
developed in Ref.~\cite{ciach:00:0,ciach:03:1,patsahan:07:0}, oscillatory decay of correlations 
was obtained on the MF level~\cite{ciach:03:1,patsahan:07:0}. This theory is a kind of simplistic DFT, combined with the
field-theoretic way of incorporating fluctuations beyond its MF version. 
The finite size of the 
cores of ions is taken into account in the calculations of the electrostatic energy,
i.e. the contributions from the overlapping cores 
are not included. As the decay of the charge density near a wall mimics the decay of the charge-density correlations in the bulk, 
one can expect that in this MF theory the oscillatory decay of the charge density should be obtained. 
Such an expectation is explicitly written 
at the end of Ref.\cite{ciach:03:1}. Unfortunately, the correlation functions were obtained in the Fourier representation, and in 
the case of the broken translational symmetry the real-space representation is required to obtain the charge-density profile.

In this work we propose an approximate version of the theory developed in 
Ref.\cite{ciach:00:0} that can be applied to
confined systems. We develop a functional of the charge- and number density of ions that has a very simple form. 
In the lowest-order approximation, the functional depends only on the local charge density, and has a form
of a single integral in real space. For this functional we
 obtain the Euler-Lagrange (EL) equation for the charge density. This EL equation
can be easily 
solved analytically, and  the theory can be systematically improved. 
Its advantage is a simplicity and possibility of obtaining approximate analytical expressions for 
the charge profile near the charged wall.

In sec.2 we shortly summar ${\cal A},\theta$,ize the theory developed for ionic systems in the bulk. 
In sec.3 we develop and solve the EL equation. Discussion and conclusions are included in sec.4. 
\section{Short summary of the theory for the bulk system}
 We consider a mixture of spherical cations with a diameter $\sigma_+$ and a charge $e_+$, and spherical
anions with a diameter $\sigma_-$ and a charge $e_-=-|e_-|$ 
dissolved in a structureless solvent with the dielectric constant
$\epsilon$. The local dimensionless number densities of the cations and the anions are denoted
by $\rho_+({\bf r})$ and 
$\rho_-({\bf r})$, respectively.
For clarity of presentation we restrict ourselves to the anion-cation symmetric case, with 
$\sigma_+=\sigma_-=\sigma$ and $e_+=|e_-|=e$, where $e$ is the elementary charge.
In general, the symmetrical case corresponds to the 
Restricted Primitive Model (RPM) with addition of the specific short-range (SR) interactions~\cite{ciach:01:1}.  In the RPM,
the interactions consist of the steric repulsion for overlapping spherical 
cores and of Coulomb interactions between 
point charges in the centers of the cores. 
The more general case of size- and charge asymmetric ions is described in Ref.\cite{ciach:05:0,ciach:07:0}.

We introduce local  dimens ${\cal A},\theta$,ionless charge and number densities at ${\bf r}$ by
\begin{equation}
 \phi({\bf r})=\rho_+({\bf r})-\rho_-({\bf r})
\end{equation}
and 
\begin{equation}
 \rho({\bf r})=\rho_+({\bf r})+\rho_-({\bf r}).
\end{equation}
In the bulk homogeneous phase $\phi=0$ and $\rho({\bf r})=\rho_0$.  

The internal energy consists of the electrostatic energy, $U_{RPM}$, and the energy associated with the SR interactions,
$U_{SR}$, $U=U_{RPM}+U_{SR}$, with
\begin{equation}
\label{URPM}
 U_{RPM}[\phi]=\frac{1}{2}\int d{\bf r}_1\int d{\bf r}_2\phi({\bf r}_1)V_C(r)g(r)\phi({\bf r}_2),
\end{equation}
where $r=|{\bf r}_1-{\bf r}_2|$, $V_C(r)$ is the Coulomb potential
and $g(r)$ is the pair distribution function.
A very rough approximation for the pair distribution function is
 \begin{equation}
 \label{g}
  g(r)=\theta(r-1).
 \end{equation}
 In (\ref{g}) and in the following,  length is measured in $\sigma$  units
(i.e. we put $\sigma\equiv 1$). In the above approximation,
$g(r)$ vanishes for $r<1$ as it should, and is equal to unity for  $r>1$, i.e. it
has a proper behavior for large $r$. 
It is convenient to introduce the dimensionless  function
\begin{equation}
\label{V*}
 V^*(r)=\frac{\epsilon\sigma}{e^2}V_C(r)g(r)=\frac{\theta(r-1)}{ r}
\end{equation}
 and the  dimensionless temperature
\begin{equation}
\label{T*}
 T^*=\frac{1}{\beta^*}=\frac{\epsilon\sigma}{e^2}k_BT,
\end{equation}
where $k_B$ and $\epsilon$ are the Boltzmann and the dielectric constant, respectively.
With the above definitions,  $\beta V_C(r)g(r)=\beta^* V^*(r)$.

In Fourier representation $V^*$ has the form
 \begin{equation}
 \label{V*k}
  \tilde V^*(k)=\frac{4\pi \cos(k)}{k^2}.
 \end{equation}
 Note that for $k>\pi/2$ we have $\tilde V^*(k)<0$, and  the charge waves with such wavenumbers, $\phi(z)\propto \cos(kz)$,
 are energetically favoured 
 compared to $\phi=0$. $\tilde V^*(k)$ assumes a minimum for $k=k_b\approx 2.46$ in  $\sigma^{-1}$-units, and the wavelength 
 of the most probable charge-density waves is $2\pi/k_b\approx 2.5$ (in $\sigma$-units).
The length scale of the charge oscillation is determined by the 
 size of the impenetrable cores.  In our theory, the contribution to the electrostatic energy coming from overlapping cores
 is not included because of the $\theta$-function present in Eq.(\ref{V*}) (see also Eq.(\ref{URPM})).
 
 For the symmetrical case, the contribution to the internal
energy associated with the SR interactions can be written in the form
\begin{equation}
 U_{SR}[\phi]=\frac{1}{2}\int d{\bf r}_1\int d{\bf r}_2\phi({\bf r}_1)K_{\phi}(r)\phi({\bf r}_2)
 +\frac{1}{2}\int d{\bf r}_1\int d{\bf r}_2\rho({\bf r}_1)K_{\rho}(r)\rho({\bf r}_2)
\end{equation}
where $K_{\phi}$ and $K_{\rho}$ denote the product of the interaction potential and the pair distribution function.
  We introduce dimensionless SR potentials by 
$\beta^* K_{\phi}^*=\beta K_{\phi}$ and $\beta^* K_{\rho}^*=\beta K_{\rho}$, with $\beta^*$ defined in (\ref{T*}).

 Let us consider  the excess grand potential
associated with particular local inhomogeneities, $\phi({\bf r})$ and  $\eta({\bf r})=\rho({\bf r})-\rho_0$,
\begin{equation}
\label{DOmega}
 \Delta\Omega[\phi,\eta]=\Omega[\phi,\rho_0+\eta]-\Omega[0,\rho_0],
\end{equation}
where 
\begin{equation}
\label{Omega}
 \Omega[\phi,\rho]=U[\phi,\rho]-TS[\phi,\rho]-\mu\int d{\bf r} \rho({\bf r}),
\end{equation}
with $U,S,T,\mu$ denoting the internal energy, entropy, temperature and the chemical potential respectively. 
 When the grand potential assumes a minimum for $\phi=0$ and $\rho=\rho_0$, then the first functional derivatives
 of $\Omega$ vanish, and terms  linear in $\phi$ and $\eta$ are absent in $\beta \Delta\Omega[\phi,\eta]$. 
 The excess grand potential takes the form
 \begin{equation}
 \label{DOm}
   \beta \Delta\Omega[\phi,\eta]= \beta \Delta\Omega_G[\phi,\eta]+
  \beta \Delta\Omega_{ho}[\phi,\eta]
 \end{equation}
 where the part containing the second-order terms in $\phi$ and $\eta$ is
 \begin{equation}
 \label{DO}
  \beta \Delta\Omega_G[\phi,\eta]=
  \frac{1}{2}\int d{\bf k} \tilde\phi({\bf k}) \tilde C_{\phi\phi}^0(k)\tilde\phi(-{\bf k})
  + \frac{1}{2}\int d{\bf k} \tilde\eta({\bf k}) \tilde C_{\eta\eta}^0(k)\tilde\eta(-{\bf k})
 \end{equation}
with
\begin{equation}
\label{phitad}
\tilde C_{\phi\phi}^0( k)= 
\rho_0^{-1}+\beta^*\big(\tilde V^*( k)+
\tilde K_{\phi}^*( k)\big),
\end{equation}
%
\begin{equation}
\label{Cetad}
\tilde C_{\eta\eta}^0( k)= \gamma_{0,2} 
+\beta^* \tilde K_{\rho}^*( k).
\end{equation}
We use Fourier representation for this part of $\Delta \Omega$ for convenience. 
The terms of higher-order in $\phi$ and $\eta$ are included in the local term
 \begin{equation}
 \label{ho}
  \beta \Delta\Omega_{ho}[\phi,\eta]=
  \sum_{n,m}^{'}\frac{\gamma_{2m,n}}{(2m)!n!} \int d{\bf r}\phi^{2m}({\bf r})\eta^n({\bf r}).
 \end{equation}
In the summation in (\ref{ho}), only terms with $2m+n>2$ are included, and 
 \begin{equation}
  \gamma_{2m,n}=\beta\frac{\partial^{2m+n}f_h}{\partial \phi^{2m}\partial \rho^n}|_{\phi=0,\rho=\rho_0}.
 \end{equation}
 
 To obtain Eqs.(\ref{DO})-(\ref{ho}) from (\ref{Omega}), we have approximated  $-T\Delta S$ by the 
 excess free energy of the hard-spheres 
 mixture in the local density approximation. 
 For the free-energy density of hard spheres, $f_h(\phi,\rho)$, we may assume the Percus-Yevick or
 Carnahan-Starling (CS) approximation. 

 Note that in (\ref{DOmega}), the excess of the grand potential associated with a particular 
 local deviation form the average values of
 $\phi$ and $\eta$ is considered. In the presence of thermal fluctuations, all such deviations
 must be taken into account with a proper weight to obtain the true grand potential.
 In such a case
 the charge-charge correlation function is defined by,
 \begin{equation}
 \label{G}
  \tilde G_{\phi\phi}( k)\equiv\langle \tilde\phi({\bf k})\tilde\phi(-{\bf k})\rangle=
  \frac{\int D\phi D\eta \tilde\phi({\bf k})\tilde\phi(-{\bf k})e^{-\beta\Delta\Omega[\phi,\eta]}}
  {\int D\phi D\eta e^{-\beta\Delta\Omega[\phi,\eta]}}.
 \end{equation}
It reduces just to 
\begin{equation}
 \tilde G_{\phi\phi}^0( k)=1/\tilde C_{\phi\phi}^0( k)
\end{equation}
when $\beta \Delta\Omega_{ho}[\phi,\eta]$ is neglected in (\ref{DOm}).
The density-density correlation function $G_{\eta\eta}$ is
defined in an analogous way, and when $\beta \Delta\Omega_{ho}[\phi,\eta]$ is neglected in (\ref{DOm}), 
it reduces to $\tilde G^0_{\eta\eta}=1/\tilde C^0_{\eta\eta}$.

In the rest of this work we limit ourselves to the pure RPM  for clarity. Generalization to the RPM+SR system is straightforward.
In the RPM, $\tilde C_{\phi\phi}^0( k)$ is given by (\ref{phitad}) with $\tilde K_{\phi}^*( k)=0$. 
In real-space representation, $G_{\phi\phi}^0(r)$ exhibits
exponentialy damped oscillatory behavior~\cite{ciach:03:1},
\begin{equation}
 rG_{\phi\phi}^0(r)=-A_{\phi\phi}\sin(\alpha_1 r+\theta)e^{-\alpha_0r},
\end{equation}
for $S_{\lambda}<T^*/\rho_0<S_K$, where $T^*/\rho_0=S_K$ and  $T^*/\rho_0=S_{\lambda}$
correspond to the Kirkwood and the $\lambda$-line respectively. For $T^*/\rho_0>S_K$, a monotonic decay of correlations occurs. The larger
one of the two decay lengths approaches the Debye length for $T^*/\rho_0\to \infty$.
For
$T^*/\rho_0<S_{\lambda}$, the correlation length is infinite, i.e. a charge-ordered phase with 
an oscillatory charge density is found in this approximation. The inverse decay length and the wave number of oscillations,
$\alpha_0$ and $\alpha_1$, are given in Ref.\cite{ciach:03:1}. 

When  $\beta \Delta\Omega_{ho}[\phi,\eta]$ is taken into account in (\ref{G}), 
then the correlation function defined in (\ref{G}) differs from $G_{\phi\phi}^0$. 
In order to obtain a better approximation for $G_{\phi\phi}$, we first note that
 in the RPM, the dependence on $\eta({\bf r})$
is strictly local. Thus, $\Delta\Omega$ can be easily minimized with respect to $\eta({\bf r})$ for each $\phi({\bf r})$.
As a result, we obtain 
\begin{equation}
\label{etaWF}
\eta(\phi({\bf r}))=\sum_{n=1}^{\infty}\frac{a_n}{n!}\phi({\bf r})^{2n} ,
\end{equation}
where $a_n$ are functions of $\rho_0$ \cite{ciach:06:2,patsahan:07:0}. 
In Ref.\cite{patsahan:07:0}
the expansion in (\ref{etaWF}) was truncated at $n=2$. From (\ref{etaWF}) and (\ref{DOm})-(\ref{ho}) 
we obtain the following approximation
for the excess grand potential
\begin{eqnarray}
\label{phi6t}
\beta\Delta\Omega[\phi]=
\frac{1}{2}\int \frac{d{\bf k}}{(2\pi)^d}\tilde\phi({\bf k})
\tilde C_{\phi\phi}^0(k)\tilde\phi(-{\bf k})
+\frac{{\cal A}_{4}}{4!}
\int d{\bf r}\phi^{4}({\bf r})+ \frac{{\cal A}_{6}}{6!}
\int d{\bf r}\phi^{6}({\bf r})+O(\phi^8),
\end{eqnarray}
 where  ${\cal A}_4$ and ${\cal A}_6$  are combinations of $\gamma_{2m,n}$ (i.e. depend on $\rho_0$) and 
   their explicit forms for the CS approximation for $f_h$ are given in Ref.\cite{ciach:06:2}.

In Ref.\cite{patsahan:07:0}  $\tilde G_{\phi\phi}( k)$ was calculated 
in  the self-consistent 
one-loop approximation. In this approximation, the $k$-dependence of $\tilde G_{\phi\phi}( k)$ is the same as the  
$k$-dependence of $\tilde G_{\phi\phi}^0( k)$, and
%
\begin{equation}
\label{C_rH}
\tilde G^{-1}_{\phi\phi}(k)=\tilde C_{\phi\phi}(k)=c_0+\beta^*\tilde V^*(k),
\end{equation}
where 
\begin{equation}
\label{C_r}
 c_0=\rho_0^{-1}+\Big(\frac{{\cal A}_4{\cal G}(c_0)}{2}+
\frac{{\cal A}_6{\cal G}(c_0)^2}{8}\Big),
\end{equation}
 and the fluctuation-induced correction depends on 
\begin{equation}
\label{calG0}
 {\cal G}(c_0)\equiv\langle \phi({\bf r})^2\rangle=
\int\frac{d{\bf k}}{(2\pi)^d} \tilde G_{\phi\phi}(k).
\end{equation}
The integral in (\ref{calG0})  diverges because of the integrand
behavior for $k\to\infty$. However, the contribution from $k\to\infty$
is unphysical (overlapping hard cores).  When the fluctuations with
$k\approx k_b$ dominate, which is the case for $T^*/\rho_0\ll S_0\approx 2.6$~\cite{patsahan:07:0}, 
then the dominant contribution to
the cutoff-regularized integral in (\ref{calG0})  comes from $k\approx k_b$ and is cutoff-independent.
In this case 
$
{\cal G}(c_0)\propto 1/\sqrt{c_0}
$\cite{brazovskii:75:0,patsahan:07:0}.
The self-consistent solution of (\ref{C_rH})-(\ref{calG0}) gives $c_0$ that 
depends on the thermodynamic state in a rather nontrivial way~\cite{patsahan:07:0}, and leads to
$\tilde C_{\phi\phi}(k_b)>0$ for $T>0$, 
i.e. the instability with respect to periodic ordering, $\tilde C_{\phi\phi}(k_b)=0$, is shifted down to $T=0$.

 \section{Euler-Lagrange equation for a confined system}
 
 In the previous section we considered the excess grand potential associated with local deviations 
 of the charge- and number density of ions from the 
 equilibrium values. 
In confinement, the inhomogeneities are induced by the presence of the boundaries. The difference between the
grand potential in a presence of the boundary and the grand potential of the same system in the bulk,
$ \Delta\Omega$, consists of two terms~\cite{evans:90:0},
\begin{equation}
\label{DO1}
  \Delta\Omega= \Delta\Omega_b+ \Delta\Omega_s.
\end{equation}
The first term in (\ref{DO1}) is proportional to the system volume and is associated with the
presence of inhomogeneities.
This contribution was considered in the previous section.
The second term contains the contribution to the grand potential associated with the fact that
the fluid beyond the system boundary is replaced by the solid wall, and the interactions with the
fluid particles are replaced by
the interactions with the particles of the wall. 

We assume that in the case of the RPM, $\Delta\Omega_b$ can be
 approximated by the expressions developed in sec.2. Moreover, we limit ourselves to
 small surface charge density that leads to small
 $\phi({\bf r})$. In such a case terms of order  $\phi^4$ can be neglected compared to terms of order
 $\phi^2$ in (\ref{phi6t}). Consistent with the assumption
 of small surface charge, we
 neglect the second and the third term in Eq.(\ref{phi6t}). However,
for the second functional 
derivative of $ \Delta\Omega_b$ we assume the
function (\ref{C_rH}), and obtain the approximation
\begin{equation}
 \label{DOr}
  \beta \Delta\Omega_b[\phi]\approx
  \frac{1}{2}\int d{\bf k} \tilde\phi({\bf k})\Big( c_0+\beta^*\tilde V^*(k)\Big)\tilde\phi(-{\bf k}).
 \end{equation}
By replacing $1/\rho_0$  by $c_0$ (compare (\ref{phitad}) and (\ref{C_rH})), 
we approximately take into account the effect of fluctuations. In Ref.\cite{patsahan:07:0} it was shown that
for large values of $T^*/\rho_0$, the parameter  $c_0$ differs from $1/\rho_0$ rather weakly.
For small values of
 $T^*/\rho_0$, however,  $c_0$ is significantly different from $1/\rho_0$, and  
 the artifact of the MF approximation - namely the continuous transition to the 
 charge-ordered phase present in MF and absent  in reality,
 is replaced by the first order transition to an ionic crystal. 
 The latter transition occurs for much higher density of ions~\cite{ciach:06:2,patsahan:07:0}.
Here we treat  $c_0$ as a parameter depending on the thermodynamic state, 
such that 
\begin{equation}
 R^*\equiv T^*c_0+\tilde V^*(k_b)>0.
\end{equation}
$R^*\to 0$ for $T^*\to 0$, and  increases for increasing $T^*$ and/or
decreasing $\rho_0$. 
In this approximation $\tilde G_{\phi\phi}(k)$ depends on the thermodynamic state through 
the single parameter  $R^*$. 
Precise dependence of $R^*$ on the thermodynamic state is not of primary 
importance in our study; we only need to ensure that  the system is in the liquid phase.
In Ref.\cite{patsahan:07:0} it was found  at the same level of approximation as in the present study that at the liquid-solid coexistence
 $R^*\approx 0.2$ for $0.25<\rho_0<0.8$.
We shall thus assume that for RTILs and RTILs mixtures
with neutral solvents the relevant values of 
$R^*$ are $0.2<R^*<0.5$.

To determine $\phi({\bf r})$ corresponding
to the minimum of $\Delta\Omega$, we shall first transform the approximate expression for 
$\Delta\Omega_b$ (Eq.(\ref{DOr})) to a more convenient form.
Our aim is to develop an expression for
$\Delta\Omega_b$ in terms of a single integral in the real-space representation.
From the minimum condition for such a functional one can obtain the Euler-Lagrange (EL) 
equation for $\phi({\bf r})$ 
that can be solved easily.

Because of the long-range of the Coulomb potential, $\tilde V^*(k)$ diverges for $k\to 0$, 
and cannot be expanded in a Taylor series. 
We note, however that the
main contribution to the electrostatic energy comes from the charge-density waves, with the wavelength 
$2\pi/k_b$
corresponding to the minimum of  $\tilde V^*(k)$ at $k=k_b$. Taking into account that  
$\tilde V^*(k)$ is an even function of $k$ (see (\ref{V*k})),
we can consider a function of $k^2$, 
and expand it about the minimum at $k^2=k_b^2$,
\begin{equation}
\label{vex}
 \tilde v^*(k^2)\equiv \tilde V^*(k)=\tilde V^*(k_b)+\sum_{n=2}^{\infty} \frac{v^{(n)}}{n!}(k^2-k_b^2)^n,
\end{equation}
where $v^{(n)}$ denotes the $n$-th derivative of $\tilde v^*(k^2)$ with respect to $k^2$ at $k^2=k^2_b$.
Using (\ref{vex}) and the equality
\begin{equation}
\label{fk}
 \int \frac{d{\bf k}}{(2\pi)^d} k^2\tilde\phi({\bf k})\tilde\phi(-{\bf k})=
 -\int d{\bf r} \phi({\bf r})\nabla^2 \phi({\bf r}),
\end{equation}
where $\nabla^2$ denotes the Laplacian, we can write (\ref{DOr}) in the form
\begin{equation}
\label{Doc}
  \beta \Delta\Omega_b[\phi]=\frac{\beta^*}{2}\int d{\bf r}\Bigg[R^*\phi({\bf r})^2
  +\phi({\bf r})\sum_{n=2}^{\infty} \frac{v^{(n)}}{n!} (-\nabla^2 -k_b^2)^n\phi({\bf r})\Bigg].
\end{equation}
%

It is instructive to consider (\ref{Doc})
with the expansion truncated at the lowest-order term, and perform integration by parts. 
Neglecting the surface terms, we obtain
\begin{equation}
\label{simple}
 \beta \Delta\Omega_b[\phi]=\int d{\bf r} \Bigg[A_0 \phi({\bf r})^2-A_2
 \Big(\nabla\phi({\bf r})
 \Big)^2+
 A_4\Big(\nabla^2\phi({\bf r})\Big)^2\Bigg]
 +...
\end{equation}
where the coefficients
\begin{equation}
 A_0=\frac{\beta^*}{2}\Big(R^*+\frac{v^{(2)}}{2}k_b^4
 \Big),
\end{equation}
\begin{equation}
 A_2=\frac{\beta^*v^{(2)}k_b^2}{2}
\end{equation}
and
\begin{equation}
  A_4=\frac{\beta^*v^{(2)}}{4}
\end{equation}
are all positive. Note that the inhomogeneities ($\nabla\phi({\bf r})\ne 0$)
are favoured by the second term in (\ref{simple}). The functional (\ref{simple})
has the form first 
proposed by Brazovskii~\cite{brazovskii:75:0} for description of systems with mesoscopic inhomogeneities. 
We should note, however that in (\ref{simple}) the first term is given by a particularly simple expression,
whereas  in general 
the term describing the local dependence on the order parameter (OP) is given by a function $f(\phi({\bf r}))$ that may have different
forms in different inhomogeneous systems. 
The Landau-Brazovskii (LB) functional was used or developed for description of block copolymers, microemulsion, 
systems with short range attraction 
and long-range repulsion (SALR), thin magnetic films with competing ferromagnetic and dipolar interactions, 
and even critical mixture with antagonistic salt
~\cite{leibler:80:0,gompper:94:0,ciach:13:0,barci:13:0,pousaneh:14:1}.

 For a semiinfinite system
with the planar confining surface at $z=0$ (in the coordinate frame where ${\bf r}=(x,y,z)$),
the EL equation for (\ref{simple}) takes the very simple form
\begin{equation}
\label{ELsimple}
 A_4\frac{d^4\Phi(z)}{dz^4}+A_2\frac{d^2\Phi(z)}{dz^2}+A_0\Phi(z)=0,
\end{equation}
where
\begin{equation}
 \Phi(z)=\int dx\int dy \phi(x,y,z)/A_S.
\end{equation}
In the above definition of the average charge at the distance $z$ from the wall,  the integration is 
over the surface in the $(x,y)$ plane with the area $A_S\to\infty$.

The solution of (\ref{ELsimple}) is a linear combination of four terms of the type 
$\exp(\lambda z)$, with $\lambda$ satisfying
the equation
\begin{equation}
\label{lambda_sol1}
 A_4\lambda^4+A_2\lambda^2+A_0=0.
\end{equation}
The solutions are
\begin{equation}
 \lambda=\pm\alpha_0\pm i \alpha_1,
\end{equation}
where 
\begin{equation}
\label{alpha0}
 \alpha_0=k_b\Bigg[
 \frac{\sqrt A -1}{2}
 \Bigg]^{1/2}
\end{equation}
\begin{equation}
\label{alpha1}
 \alpha_1=k_b\Bigg[
 \frac{\sqrt A +1}{2}
 \Bigg]^{1/2}
\end{equation}
\begin{equation}
\label{A}
 A=1+\frac{4R^*}{|\tilde V^*(k_b)|(1+k_b^2/2)}.
\end{equation}
\begin{figure}[h]
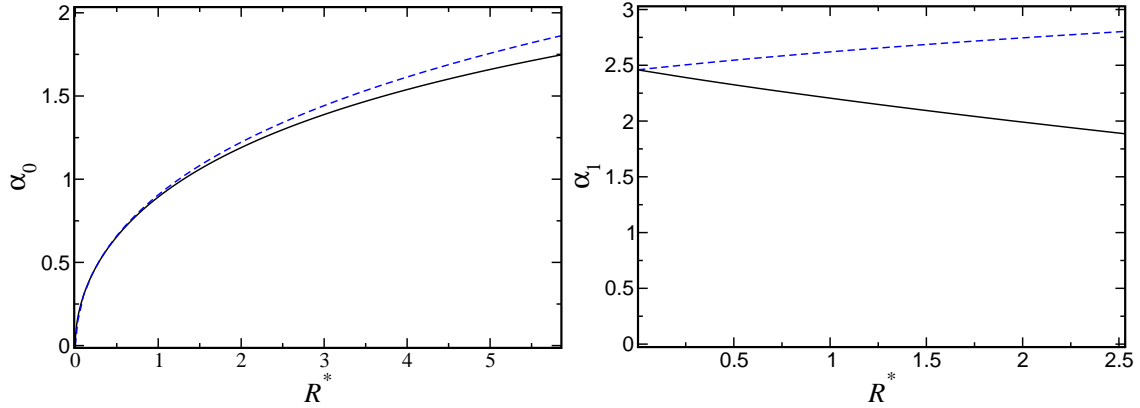

 \centering
 \includegraphics[scale=0.3]{fig1a.eps}
  \includegraphics[scale=0.3]{fig1b.eps}
\caption{The inverse decay length $\alpha_0$ (left),
 and the wavenumber $\alpha_1$ (right) 
 as  functions of $R^*=T^*c_0+\tilde V(k_b)$. 
 $R^*$ is an increasing function of the dimensionless temperature
 $T^*$ and a decreasing function of the dimensionless number density $\rho_0$. 
 Dashed lines represent Eqs.(\ref{alpha0}) and (\ref{alpha1}).
 Solid  lines represent the corresponding inverse lengths in the expression for the
 correlation function obtained in Ref.\cite{ciach:03:1}.
}
\label{alphy}
\end{figure}
In the semiinfinite system $\Phi(z)\to 0$ for $z\to \infty$, and 
\begin{equation}
\label{phiz}
 \Phi(z)={\cal A}e^{-\alpha_0 z}\sin(\alpha_1 z+\theta),
\end{equation}
where the amplitude ${\cal A}$ and the phase $\theta$ should be determined from
the boundary conditions. The latter should follow from the surface contributions to $\Delta\Omega$.

 When we apply the continuous description to the structure at the molecular scale,
we face some  ambiguities, or dependence on the definitions, leading to problems with finding a unique 
expression for the second term in (\ref{DO1}). In particular, it is necessary to assume where the
system boundary ($z=0$) and the surface charges are precisely
located.
Here we assume that  the charge is uniformly distributed over the core of the particle.
In order to precisely define $\Phi(z_0)$, we consider the slab parallel to the wall,
$z\in [z_0-1/2,z_0+1/2]$ (recall that length is in $\sigma$-units).
We assume that $\frac{\pi}{6} \Phi(z)$ denotes the fraction of the volume  of this slab that is occupied by 
the cations minus the fraction of the volume  of this slab that is occupied by 
the anions. 

\begin{figure}[h]
 \centering
 \includegraphics[scale=0.3]{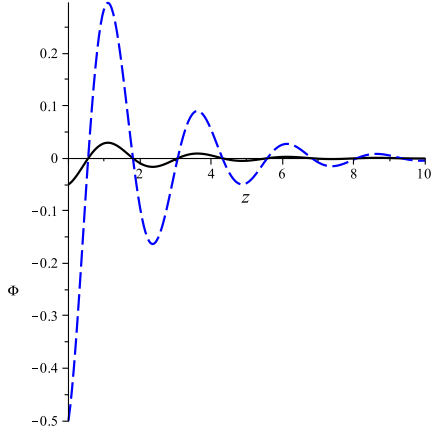}
\caption{The  charge-density profile $\Phi(z)$ ((\ref{phiz}) and (\ref{calA})) 
(in $e\sigma^{-3}$-units),  for $R^*=0.25$.
The surface charge density (in $e\sigma^{-2}$-units) is $\sigma_0=-0.05$ (solid line) and $\sigma_0=-0.5$ 
 (dashed line). Distance is in units of the ion diameter $\sigma$. When $\sigma=1nm$,
the two values of the surface charge density correspond to 
$\sigma_0\approx 0.8\mu C/cm^2$ and $\sigma_0\approx 8\mu C/cm^2$ 
respectively, the case studied in simulation in Ref.\cite{fedorov:08:0}.}
\label{phi}
\end{figure}

We choose simulations of Ref.\cite{fedorov:08:0} for comparison with predictions of our theory,
because these simulations concern the RPM, i.e. the model to which the current version of our theory
is restricted.
In order to compare our results with the results of the simulations in Ref.\cite{fedorov:08:0},
we assume that the system boundary consists of homogeneously charged spheres,
with the centers homogeneously distributed  at the $z=0$ plane.
The total charge of these fixed spheres per
 area of the $z=0$ plane is equal to $\sigma_0$. 
\begin{figure}[h]
  \centering
 \includegraphics[scale=0.25]{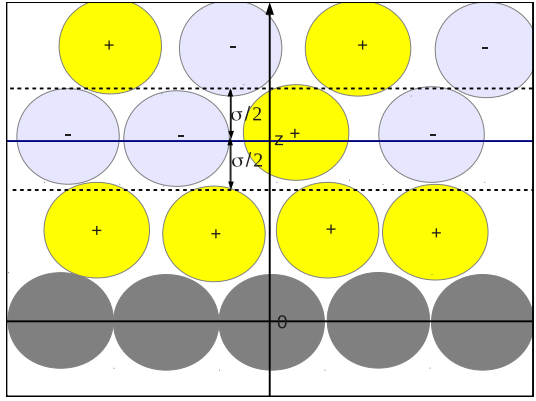}
 \caption{Cartoon illustrating  $ \Phi(z)$ in this continuous theory applied to the microscopic length-scale. 
 $\frac{\pi}{6} \Phi(z)$ is the fraction 
 of the volume of the slab between the dashed lines at $z-\sigma/2$ and $z+\sigma/2$ that is occupied by the 
 cations (plus sign) minus the fraction of the volume  of the slab that is occupied by 
the anions (minus sign). The system boundary consists of negatively charged dark hard spheres with the centers 
fixed at the $z=0$ plane.
 We include the charge of the boundary layer in $\Phi$, hence $ \Phi(0)=\sigma_0$, where $e\sigma_0$ is 
 the total charge of the surface layer divided by
the area $A_S$ of the $z=0$-plane.
 }
 \label{cartoon}
\end{figure}
Including this
fixed charge in the charge profile $\Phi(z)$, we obtain the condition $\Phi(0)=\sigma_0$. 
Moreover, we require the electroneutrality of the whole system,
$\int_0^{\infty}dz\Phi(z)=0$.
 From the two above conditions we
 obtain 
\begin{equation}
\label{calA}
{\cal A}=\frac{\sigma_0}{\sin \theta},
\hskip1cm  \theta=\arctan\Big(-
\frac{\alpha_1}{\alpha_0}
\Big).
\end{equation}
Two  examples of the charge-density profile  are shown in Fig.\ref{phi}. We choose small surface charges,
since the approximate theory is valid only for small $\Phi$.

From the charge profile we can obtain the electrostatic potential
$\Psi(z)$, using the Poisson equation. For our geometry and for $\Psi(z)\to 0$ for $z\to\infty$,
we can write the Poisson equation in the form (recall that $\Phi$ and $z$ are dimensionless)
\begin{equation}
\label{Psi}
\Psi(z)=\frac{4\pi e}{\epsilon\sigma}\int_z^{\infty} dz'(z-z')\Phi(z').
\end{equation}
The explicit expression can be easily obtained from (\ref{Psi}), (\ref{phiz}) and (\ref{calA}),
\begin{equation}
\label{Psiexplicit}
\Psi(z)=\frac{4\pi e\sigma_0}{\epsilon\sigma\sin \theta(\alpha_0^2+\alpha_1^2)^2}
\Big[\Big(\alpha_1^2-\alpha_0^2\Big)\sin(\alpha_1 z+\theta)
-2\alpha_0\alpha_1\cos(\alpha_1 z+\theta)\Big]e^{-\alpha_0 z}.
\end{equation}

\begin{figure}[h]
 \centering
 \includegraphics[scale=0.3]{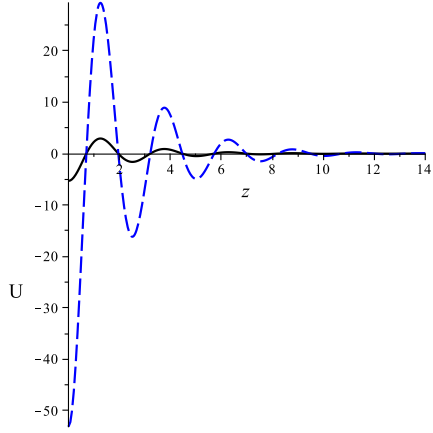}
\caption{The  re-scaled electrostatic potential  $U=\Psi(z)e/k_BT$ (dimensionless),
for $R^*=0.25$, $\epsilon=1$ and $T=300K$.
The surface charge density (in $e\sigma^{-2}$-units) is $\sigma_0=-0.05$ (solid line) and $\sigma_0=-0.5$ 
 (dashed line). Distance is in units of the ion diameter $\sigma$. When $\sigma=1nm$,
the two values of the surface charge density correspond to 
$\sigma_0\approx 0.8\mu C/cm^2$ and $\sigma_0\approx 8\mu C/cm^2$ 
respectively, the case studied in simulations in Ref.\cite{fedorov:08:0}.}
\label{Psifig}
\end{figure}

In Fig.\ref{Psifig} we plot $U(z)=\Psi(z)e/k_BT$  for ions with $\sigma=1nm$, $\epsilon=1$, 
and surface charge density $\sigma_0\approx 0.8\mu C/cm^2$ and $\sigma_0\approx 8\mu C/cm^2$,
in order to compare the order of magnitude of our result
with the simulations performed in Ref.\cite{fedorov:08:0}. 

\begin{figure}[h]
 \centering
 \includegraphics[scale=0.3]{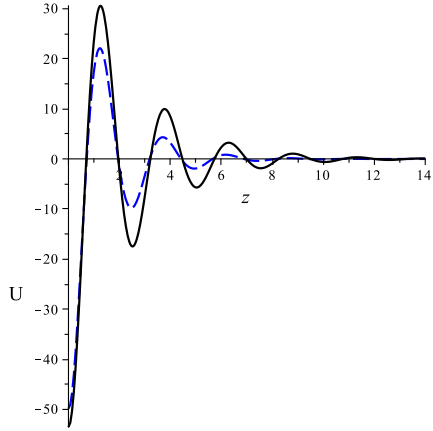}
\caption{The  rescaled electrostatic potential  $U=\Psi(z)e/k_BT$ (dimensionless)
for the surface charge density $\sigma_0=-0.5$ (in $e\sigma^{-2}$-units), $\epsilon=1$
and $T=300K$. When $\sigma=1nm$,
 the surface charge density corresponds to $\sigma_0\approx 8\mu C/cm^2$.
 Solid and dashed lines correspond to $R^*=0.22$  
and  $R^*=0.5$ respectively.
 Distance is in units of the ion diameter $\sigma$. }
\label{Psifig1}
\end{figure}

We did not try to precisely adjust the thermodynamic state to the
thermodynamic state in the simulations. Instead,
 we compared the charge and potential profiles 
for different values of $R^*$ within the stability region of the liquid phase (Fig.\ref{Psifig1}).
For $R^*\le 0.5$, the potential 
has the shape very similar to the shape obtained in simulations already 
in the simplest version of our theory. The decay length decreases with increasing  $R^*$,
(see Fig.\ref{alphy}), but
for the relevant value of $R^*$, $0.2<R^*<0.5$, we see essentially 
the same number of peaks as in the simulations.
Moreover, the order of magnitude for the rescaled potential is in very good agreement
with the simulation results. We cannot expect quantitative agreement in such a simplified theory. Moreover,
in contrast to the simulations, we did not take into account the polarization effects of the ions.

The approximation for $\Delta\Omega$ can be improved when more terms are taken into
account in the expansion in  Eq.(\ref{Doc}). 
When $N$  terms in Eq.(\ref{Doc}) are taken into account, then the EL equation is
\begin{equation}
 R^*\phi({\bf r})
  +\sum_{n=2}^{N} \frac{v^{(n)}}{n!} (-\nabla^2 -k_b^2)^n\phi({\bf r})=0.
\end{equation}
The solution has the exponential form $\exp(\lambda z)$, where $\lambda$ satisfies the above equation
with $\nabla^2$ replaced by $\lambda^2$. 
Note, however that the series in (\ref{Doc}) converges (see (\ref{vex})),
and we can obtain $\lambda^2$ by solving
\begin{equation}
 \tilde C_{\phi\phi}(k)=c_0+\beta^*\tilde v(k^2)=0,
\end{equation}
with  $k^2$ replaced by $-\lambda^2$.
The zeros of the above function were determined in Ref.\cite{ciach:03:1}, and we compare
the so obtained $\alpha_0$ and $\alpha_1$ with the result of our simple approximation 
(\ref{alpha0})-(\ref{alpha1}) in Fig.\ref{alphy}.
 While our simple approximation (\ref{alpha0}) rather well reproduces $\alpha_0$ for $R^*<5$, $\alpha_1$
 given by (\ref{alpha1})
 is more and more overestimated for increasing $R^*$. In the  
 region of interest in ionic liquids (  $0.2\le R^*<0.5$), however, the simple approximation
 is quite reasonable; for $R^*\le 0.5$ the relative deviation is $\le 9\%$.
 The charge-density profile obtained by minimization
 of the functional (\ref{Doc}) is only slightly
 different from the profile shown in Fig.\ref{phi} and  obtained by
 the  minimization of the simple approximation (\ref{simple}).
 
The accuracy of the approximate EL (\ref{ELsimple}) is directly linked with
the accuracy of the correlation function (\ref{C_rH}) with 
$\tilde V^*(k)$ approximated by the expansion (\ref{vex}) truncated at $n=2$.
For $2<k<3$ the accuracy of the approximate form of the 
correlation function
increases when $R^*$ 
decreases to small values, consistent with the result for $\alpha_0, \alpha_1$. 
The very simple approximation is thus reasonable 
 for both the correlation function and the structure near the wall for the thermodynamic states 
 relevant for ionic liquids.
\begin{figure}[h]
 \centering
 \includegraphics[scale=0.4]{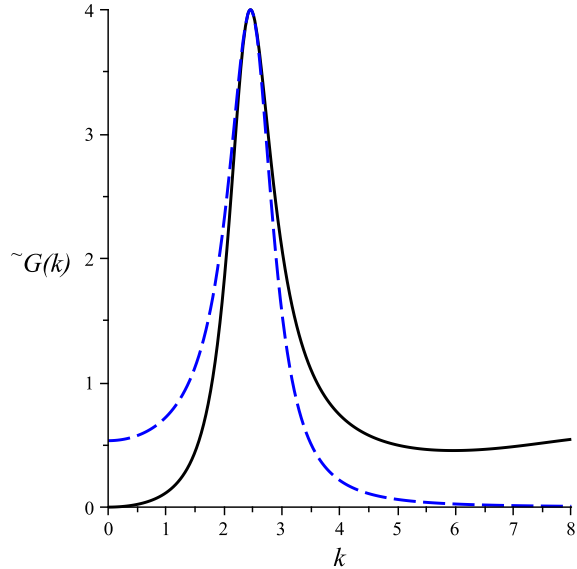}
\caption{The correlation function for $R^*= 0.25$ with $\tilde V^*(k)$  given by (\ref{V*k})  (solid line) 
and by the expansion (\ref{vex}) 
truncated at $n=2$ (dashed line).
}
\label{Gfig}
\end{figure}
\section{Summary and conclusions}

The aim of this work was a development of a simple theory for the structure of 
the ionic liquid or ionic liquid mixture with a neutral solvent, near a charged boundary. 
The theory is based on the mesoscopic theory for ionic systems developed for the bulk in 
Refs.~\cite{ciach:00:0,ciach:01:1,ciach:03:1,ciach:05:0,ciach:07:0,patsahan:07:0}.
Even though our theory is essentially of the MF character, we still obtain oscillatory decay of the
charge density for increasing distance from the charged surface. For the entropy part of the grand potential 
we have made the local density approximation, therefore packing effects of hard spheres are not 
taken into account through this contribution to the grand potential. We have not included, however, the 
contributions to the electrostatic energy associated with overlapping cores of the ions. We take into account
that only pairs of ions separated by distances larger than the sum of their radii contribute to 
the internal energy simply by assuming that the 
pair distribution function in the
expression for the internal energy vanishes for shorter distances. By not including the electrostatic energy of 
overlapping cores, we set the distance between the anion and the cation that corresponds to the minimum of the 
energy. This is the origin of the oscillatory decay of the charge density on the length scale set by the size of the ions.

Our results for the charge profile and for the electrostatic potential have been obtained from the EL equations
for the excess grand potential approximated by
 the functional of the second order in the local charge density $\phi$ (see Eq.(\ref{DOr})). In addition, we took into account fluctuations by
renormalization of the parameter depending on the thermodynamic state in this functional. However, in MF the 
functional has precisely the same form, and leads to the same shape of the charge density profile, except
that for overestimated values of $T^*/\rho_0$. 

We conclude that it is not the MF approximation that leads to the monotonic decay of the charge density and the potential
near the charged surface, but the specific version of MF, with point-like charges.
Neglecting the finite size of ions is justified at very low densities, but not in IL. 
The key factor leading to the oscillatory decay of the potential is the presence of the impenetrable cores.
In the DFT theories the sophisticated form of the entropy of the hard spheres leads to the oscillatory profile~\cite{roth:02:0}.
In our theory, it is sufficient to take into account the finite size of the ions only in the expression
for the internal energy. 

We have made other simplifying assumptions in addition to the local density approximation
to obtain a simple analytic expression for the charge density. 
Because of that, we cannot expect quantitative
agreement of our predictions with exact or simulation results. Still, the agreement with simulations
for the RPM is surprisingly good. The very simple analytical 
expressions for the charge and potential profiles is a clear advantage of the theory.
Moreover, the theory can be extended in many ways. We can include SR interactions, or consider ions 
of different sizes, as done in the bulk in Ref.\cite{ciach:07:0,patsahan:12:0}, and include higher order 
terms in (\ref{phi6t}) in the case of larger surface charge. 

In Ref.\cite{simonin:99:0}, co-authored by Lesser Blum, the first sentence in the introduction reads:
``Analytical theories of ionic solutions are of practical interest,
since the properties of real systems can be represented in terms
of (hopefully) relatively simple equations.''
We hope that the present work is in the same spirit. The properties of real systems are represented
in terms of very simple equations (\ref{phiz}) and (\ref{Psiexplicit}).

 Acknowledgement:
 
 This article is dedicated to the memory of Lesser Blum.
 
I  would like to acknowledge interesting discussions with the participants of the CONIN workshop
financed by EU MSC 734276, 
 especially with prof. Ruth Lynden-Bell and prof. Luis Varela. 
 I'm also grateful to Dr. Oksana Patsahan for comments on the manuscript. 
 This project has received funding from the European Union’s Horizon 2020 research and innovation
 programme under the Marie Skłodowska-Curie grant agreement No 734276,
and from the National Science Center grant 2015/19/B/ST3/03122.

\end{document}